\documentclass
[twocolumn,english,aps,superscriptaddress,nofootinbib]{revtex4-2}%
\usepackage{amsfonts}
\usepackage[T1]{fontenc}
\usepackage{amsmath}
\usepackage{amssymb}
\usepackage{babel}
\usepackage[percent]{overpic}
\usepackage{graphicx}
\usepackage{lastpage}
\usepackage{xcolor}
\usepackage[hidelinks]{hyperref}
\usepackage{braket}%
 
\begin{document}

\title{Effects of superradiance on relativistic Foldy-Wouthuysen densities }
\author{F. Daem}
\affiliation{Laboratoire de Physique Th\'eorique et Mod\'elisation, CNRS Unit\'e 8089, CY
	Cergy Paris Universit\'e, 95302 Cergy-Pontoise cedex, France}
\author {A.\ Matzkin}
\affiliation{Laboratoire de Physique Th\'eorique et Mod\'elisation, CNRS Unit\'e 8089, CY
	Cergy Paris Universit\'e, 95302 Cergy-Pontoise cedex, France}

\begin{abstract}
Recent interest in the studies of structured states obtained in relativistic electron beams 
has highlighted the use of two alternative descriptions, each based on a different wavefunction 
and the related  space-time  density.
Although both wavefunctions obey the Dirac equation (one directly and the other through a Foldy-Wouthuysen transformation)
they lead to different dynamics 
and properties, such as the presence or absence of spin-orbit interactions. 
In this work we investigate wavepacket dynamics for the Klein-Gordon equation, 
which displays the same ambiguity regarding the choice of different densities, in a setting involving Klein tunneling across a series 
of supercritical potential barriers. Relying on the superradiant character of this setting, we obtain 
solutions to the wavepacket dynamics
indicating that the density based on a Foldy-Wouthuysen transformation of the wavefunction
can be locally amplified outside the light-cone. In principle, the exponential increase of the charge due to
the field inhomogeneities can lead to an arbitrarily large amplification over macroscopic distances. These
results question the interpretation of the Foldy-Wouthuysen density as a fundamentally correct probability or charge density.

\end{abstract}
\maketitle

\newpage

Defining physical wavefunctions and their associated probability or charge
densities in relativistic quantum mechanics has been a long-standing issue.
The problem has resurfaced recently in connection with the production of
high-energy shaped electron beams \cite{review-eb}. These beams can be described by
the Hamiltonian $h_{D}$ of the standard (\textquotedblleft
canonical\textquotedblright) Dirac equation, characterized by the presence of
spin-orbit couplings. According to
this approach, the total current does not present vortex lines
\cite{bialy2017}, contrarily to the Schr\"{o}dinger equation solutions.\ A
strictly equivalent Hamiltonian $h_{FW}$, obtained by a Foldy-Wouthuysen (FW)
transformation \cite{FW,greinerRQM} of $h_{D}$, yields instead solutions for which the
spin and orbital angular momentum are separately conserved, leading to the
existence of electron vortices in accordance with the dynamics in the
low-energy nonrelativistic regime \cite{barnett2017,silenko2018}.

While there is no consensus \cite{bliokh2017,viewpoint,bialy-comm,silenko-comm,choi2020,oppeneer2020}
on which of the two pictures should be regarded as
the more physical one, the
culprit is well-identified: it relies on the choice of a position operator, a
problem that was pinpointed early on with the advent of relativistic quantum
mechanics \cite{pryce,newton-wigner,kalnay1971}. As such, the same issue appears not only
for other fermions described by the Dirac equation, but more
generally for the other relativistic wave equations as well
\cite{case,silenko2020}.\ Typically, the densities
given by the canonical Hamiltonian present interferences between positive and
negative energy components, transform in a covariant way, and propagate
causally; the corresponding position operator displays unusual properties (eg,
its derivative is not related to the momentum and can be hardly interpreted as
a velocity) and has no classical limit. In the FW formulation instead the
classical limit is well-defined and the velocity operator is the quantized
version of the classical relativistic velocity; but the FW densities are not
manifestly covariant, and their time evolution displays an exponentially small
leak outside the light-cone, on distances of the order of the Compton
wavelength of the particle. The non-causal character of the propagation of FW
densities can be rigorously proved mathematically \cite{hegerfeldt-prl}, but
it has been argued \cite{ruijsenaars,pavsic,lorce2022,choi2024} that the resulting
non-causality is so small that in practice it would be undetectable, therefore
leaving untouched the physical acceptability of the FW\ densities; similar
arguments have been put forward
\cite{rosenstein-usher,ruijgrok,eckstein,xabi} for the densities of the
relativistic Schr\"{o}dinger equation, directly related to the FW densities
for field-free scalar particles.

\begin{figure*}[tb]
	\centering
	\begin{overpic}[width=0.32\textwidth]{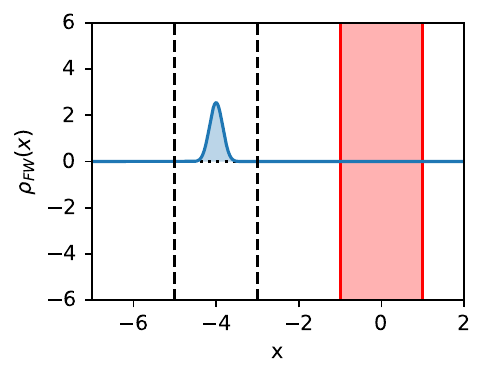}
		\put(0,70){\large\textbf{(a)}} 
	\end{overpic}
	\begin{overpic}[width=0.32\textwidth]{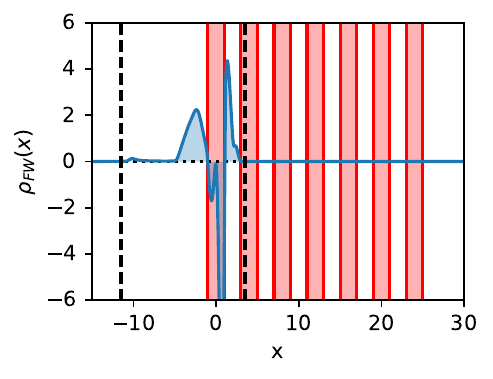}
		\put(0,70){\large\textbf{(b)}} 
	\end{overpic}
	\begin{overpic}[width=0.32\textwidth]{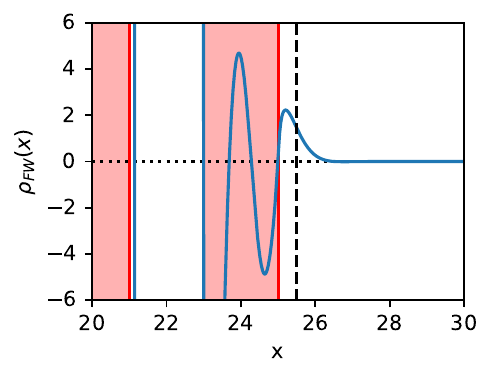}
		\put(0,70){\large\textbf{(c)}} 
	\end{overpic}
	\caption{Evolution of the charge density. (a) The initial ($t=t_0=0$) wavepacket (shown in blue) is 
		localized within a compact support delineated by the vertical dashed lines. The first potential
		barrier, centered at $x=0$ is shown in red  (we use natural units
		here and throughout: $\hbar=1$, $c=1$, $\lambda_{C}=\hbar/mc=1$).
		(b) The wavepacket is shown at $t_{b}=6.5$ along with the seven evenly spaced supercritical potential 
		barriers (in red). The position at $t=t_b$ of the light-cone emanating from the left and right edges of the initial
		compact support shown in (a) is displayed with vertical dashed lines.  (c) Snapshot at $t_{c}=28.5$ as
		the front of the charge density exits the seventh potential barrier; the dashed line gives the position of the light-cone
		emanating from the right edge of the initial compact support. 
		The parameters of each potential barrier $V_i$ are $V_{0}=5$, $L=2$ and $\epsilon=20$ [see Eq. \eqref{numv}], 
		and from the initial wavepacket we have  $\Delta_{x}=2$, $x_{0}=-4$, and $p_{0}=2$ [see Eq. \eqref{eq:initwf}].}%
	\label{fig:7potentialsEvolution}%
\end{figure*}

In this work, we will examine FW densities for a scalar boson of mass $m$
obeying the Klein-Gordon (KG) equation in the presence of a series of
supercritical potentials. Indeed, the well-known superradiant\footnote{Note that “superradiance” is used here in its standard meaning in relativistic wave dynamics, referring to the amplification of a wavefunction upon scattering from a supercritical potential. This is not directly related to other usages involving collective radiation, although both involve coherent amplification.} character of the
KG equation with a potential of strength $V_{0}>2mc^{2}$ implies that the
wavefunction amplitudes are amplified each time they cross a region of field
inhomogeneities \cite{manogue,dombey}. For a rectangular barrier, the
transmission amplitude can be computed analytically \cite{wavpackets-PRA-2021}
for each energy eigenstate that composes the wavepacket, and is typically
greater than $1$. Hence a boson, say a scalar meson, tunneling successively
through a sufficient number of of supercritical potentials would see the
outgoing transmitted wavefunction amplified exponentially relative to the
amplitude of the initial wavepacket. In the FW picture, this amplification
should also affect the otherwise exponentially small fraction of the
wavefunction that propagates outside the light cone (OLC). We  present
below numerical evidence that this is indeed the case.

The situation we are considering is pictured in Fig. \ref{fig:7potentialsEvolution}.\ A
particle wavepacket sits at $t=t_0$ in a field free zone and is launched
towards a supercritical rectangular-like barrier of width $L$. The transmitted
part of the wavepacket\ undergoes Klein tunneling and then propagates in a
field-free zone before encountering a second potential barrier, and then then a third barrier and so on. The
standard Klein-Gordon equation with a static scalar electromagnetic potential
$V(x)$ in the first-order (or so-called Schr\"odinger) form is $i\hbar\partial_{t}\Psi=\hat
{\mathcal{H}}\Psi$ with
\begin{equation}
\hat{\mathcal{H}}=mc^{2}\sigma_{3}+\frac{\hat{p}^{2}}{2m}(\sigma_{3}%
+i\sigma_{2})+V(\hat{x}) \label{eq:canonical-hamiltonian}%
\end{equation}
and $\Psi=\left(  \varphi,\chi\right)  $ being the 2-component wavefunction
\cite{greinerRQM}. $\sigma_{i}$ are the Pauli matrices, $\hat{p}$ is the
momentum operator and $\hat{x}$ is a so called ``position operator'' defined in this representation as $i\hbar\partial_p$. Given an
initial wavepacket $\Psi(t_{0},x),$ $\Psi(t,x)$ is readily computed by solving
numerically the KG equation (with $\hat{p}\rightarrow-i\hbar\partial_{x}$
one gets an ordinary differential equation that can be solved with standard
methods \cite{bauke}) or by a semi-analytical approach based on obtaining the
transmission amplitude from a multiple scattering expansion
\cite{wavpackets-PRA-2021}. The charge density $\rho(x)$ is then obtained as
$\rho(x)=q\Psi^{\dagger}(x)\sigma_{3}\Psi(x)$ -- recall that $\hat
{\mathcal{H}}$ is pseudo-Hermitian \cite{mostafazadeh} and that the KG scalar
product is a non-positive definite pseudo-Hermitian inner product defined by
$\braket{\Psi_{1} | \Psi_{2}}_{\sigma_3} =\int
dx\Psi_{1}^{\dagger}(x)\sigma_{3}\Psi_{2}(x)$.

The aim of a FW\ transformation is to uncouple the two components of the
wavefunction, $\varphi$ and $\chi$, linked to particle and antiparticle
amplitudes respectively. Given that a supercritical potential always mixes the
two components in the region of field inhomogeneities, the relevant
transformation to implement is the free one, corresponding to the ``asymptotic'' initial
and final states, both
lying in field-free regions (for definiteness we will consider
preparing and detecting a pure particle -- the quantum
state having no negative energy contribution). The exact pseudo-unitary FW transformation of the
free KG equation is well-known to be given by
\cite{hermanfeshbachElementaryRelativisticWave1958,greinerRQM}%
\begin{equation}
\hat{\mathcal{U}}(p)=\left(  4mc^{2}E_{p}\right)  ^{-1/2}\left[  (mc^{2}%
+E_{p})-(mc^{2}-E_{p})\sigma_{1}\right]  \label{eq:KG-FW}%
\end{equation}
with $E_{p}=\sqrt{p^{2}c^{2}+m^{2}c^{4}}$. The Hamitonian
\eqref{eq:canonical-hamiltonian} is then transformed as $\hat{\mathcal{H}%
}_{FW}=\hat{\mathcal{U}}\hat{\mathcal{H}}\hat{\mathcal{U}}^{-1}$ and the KG
equation becomes%
\begin{equation}
i\hbar\frac{\partial}{\partial t}\ket{\Psi_{FW}(t)}=\left(  \left(  \hat
{p}^{2}c^{2}+m^{2}c^{4}\right)  ^{1/2}\sigma_{3}+\hat{\mathcal{V}}%
_{FW}\right)  \ket{\Psi_{FW}(t)} \label{eq:schrodinger-FW}%
\end{equation}
with $\left\vert \Psi_{FW}\right\rangle =\hat{\mathcal{U}}\left\vert
\Psi\right\rangle $ and $\hat{\mathcal{V}}_{FW}=\hat{\mathcal{U}}%
\hat{\mathcal{V}}\hat{\mathcal{U}}^{-1}$. The KG\ equation in the FW picture
cannot be solved as a differential equation in the position representation due
to the presence of the square-root of a differential operator. Instead, Eq.
\eqref{eq:schrodinger-FW} is written as an integral equation in the momentum
representation: the kinetic term is trivial and the potential becomes%
\begin{equation}%
\bra{p}\hat{\mathcal{V}}_{FW}\ket{\Psi}=\frac{1}{\sqrt{2\pi\hbar}}%
\int_{-\infty}^{+\infty}\mathrm{d}p^{\prime}~\Tilde{V}(p-p^{\prime})\mathcal{U}%
(p)\mathcal{U}^{-1}(p^{\prime})\Psi(p^{\prime})
\label{poti}%
\end{equation}
with $\Tilde{V}(p)=\frac{1}{\sqrt{2\pi\hbar}}\int_{-\infty}^{+\infty
}V(x)e^{-ipx/\hbar}\mathrm{d}x$. The modified FW density is then defined as
\begin{equation}
\rho_{FW}(x)=q\Psi_{FW}^{\dagger}(x)\sigma_{3}\Psi_{FW}(x) \label{xdensity}\,.%
\end{equation}
with%
\begin{equation}
\Psi_{FW}(t,x)=\frac{1}{\sqrt{2\pi\hbar}}\int_{-\infty}^{\infty}\Psi
_{FW}(t,p)e^{ipx/\hbar}\mathrm{d}p \,. \label{position}%
\end{equation}

Let us now assume an initial wavepacket defined on a 
compact spatial support and a potential 
$V(x)=\sum_{i=1}^{N}V_{i}(x)$ consisting of $N$
identical rectangular potential barriers of width $L$ and strength $V_{0}$
placed at $x=x_{i}$,%
\begin{equation}
V_{i}(x)=V_{0}\left[  \theta(x-x_{i}+L/2)-\theta(x-x_{i}-L/2)\right] \,.
\label{rectb}%
\end{equation}
The first step is to obtain the eigenstates of $\hat{\mathcal{H}}_{FW}$. Like
$\hat{\mathcal{H}}$, $\hat{\mathcal{H}}_{FW}$ is pseudo-Hermitian and admits
a decomposition on a biorthogonal eigenbasis $\left\{
\ket{r_{\epsilon_\lambda}},\ket{l_{\epsilon_\lambda}}\right\}  $ where
$r_{\epsilon_{\lambda}}$ and $l_{\epsilon_{\lambda}}$ denote respectively the
right and left eigenvectors associated with the eigenvalue $\epsilon_{\lambda}$
\cite{mostafazadeh}, such that%
\begin{equation}
\left\{
\begin{array}
[c]{ll}%
\hat{\mathcal{H}}_{FW}\ket{r_{\epsilon_\lambda}}=\epsilon_{\lambda
}\ket{r_{\epsilon_\lambda}} & \\
\bra{l_{\epsilon_\lambda}}\hat{\mathcal{H}}_{FW}=\epsilon_{\lambda
}\bra{l_{\epsilon_\lambda}} & 
\end{array}
\right.  \label{eq:eigen-rl}%
\end{equation}
with $\braket{l_{\epsilon_\lambda}|r_{\epsilon_{\lambda'}}}=\delta
_{\epsilon_{\lambda},\epsilon_{\lambda^{\prime}}}$. We have as well
$\mathcal{H}_{FW}^{\dagger}\ket{l_{\epsilon_\lambda}}=\epsilon_{\lambda}%
^{\ast}\ket{l_{\epsilon_\lambda}}$ (the role of complex eigenvalues, which
appear in pairs, is crucial to superradiance [Appendix~\ref{supp-C}]). From the KG equation
\eqref{eq:schrodinger-FW} along with Eq. \eqref{poti}, we obtain the eigenvalue
equation for the right 2-component eigenvector%
\begin{equation}
\int_{-\infty}^{\infty}\mathrm{d}p^{\prime}\mathcal{K}(p,p^{\prime
})r_{\epsilon_{\lambda}}(p^{\prime})=\epsilon_{\lambda}r_{\epsilon_{\lambda}%
}(p) \label{eq:integral-2}%
\end{equation}
where%
\begin{equation}
\mathcal{K}(p,p^{\prime})=E_{p}\delta(p-p^{\prime})\sigma_{3}+\frac{1}%
{\sqrt{2\pi\hbar}}\Tilde{V}(p-p^{\prime})\mathcal{U}(p)\mathcal{U}%
^{-1}(p^{\prime}) \,.
\end{equation}
Eq. \eqref{eq:integral-2} is solved numerically \cite{num} for the eigenvalues and right
eigenvectors (see Appendices \ref{supp-A} and \ref{supp-B} for details). We have then all we need to compute the
evolved wavefunction in the momentum representation,%
\begin{equation}
\Psi_{FW}(t,p)=\int_{-\infty}^{\infty}e^{-i\epsilon_{\lambda}(t-t_0)/\hbar
}r_{\epsilon_{\lambda}}(p)\braket{l_{\epsilon_\lambda}|\Psi_{FW}(t_0)}\mathrm{d}%
\lambda
\label{eq:psi-FW}
\end{equation}
where the coefficients $\braket{l_{\epsilon_\lambda}|\Psi_{FW}(t_0)}$ are
computed using the \textquotedblleft left\textquotedblright\ eigenvectors
$l_{\epsilon_{\lambda}}(p)$,%
\begin{equation}
\braket{l_{\epsilon_\lambda}|\Psi_{FW}(t_0)}=\int_{-\infty}^{\infty}%
l_{\epsilon_{\lambda}}^{\ast}(p)\Psi_{FW}(t_{0},p)\mathrm{d}p \,,
\end{equation}
where $\Psi_{FW}(t_{0},p)$ is the initial wavefunction in the momentum
representation. We finally retrieve the Foldy-Wouthuysen density $\rho
_{FW}(t,x)$ with the help of Eqs. \eqref{xdensity} and \eqref{position}.

\begin{figure*}[ptb]
	\centering
	\begin{overpic}[width=0.49\textwidth]{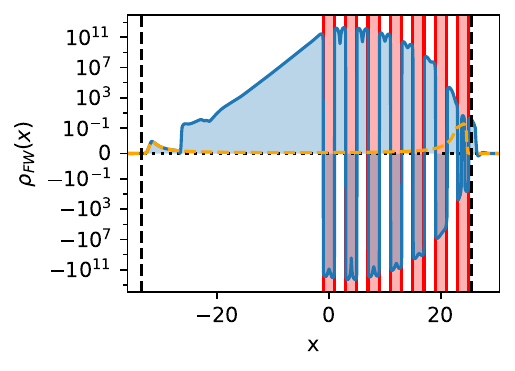}
		\put(0,70){\large\textbf{(a)}} 
	\end{overpic}
	\begin{overpic}[width=0.49\textwidth]{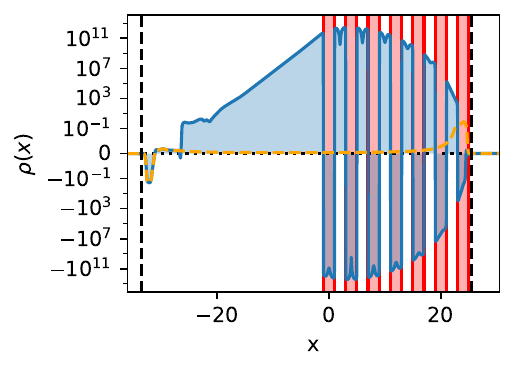}
		\put(0,70){\large\textbf{(b)}} 
	\end{overpic}
	\caption{Charge
		density in logarithmic scale (except in the range
		$[-10^{-1},10^{-1}]$, where the scale is linear) at $t=t_{c}$ [see Fig.
		\ref{fig:7potentialsEvolution}(c)], emphasizing the superradiant
		character of Klein tunneling;  parameters and units as in Fig.
		\ref{fig:7potentialsEvolution}. The charge density is plotted in blue and the
		potential barriers in red. The dashed orange line represents the 
		charge density evolved from the same initial wavepacket shown in Fig. 	\ref{fig:7potentialsEvolution}(a)
		but in the absence of any potential (free evolution).
		(a) Foldy-Wouthuysen density. (b) Standard Klein-Gordon density computed in the initial representation with the same initial density and potential barriers as in (a). 
		}%
	\label{fig:density-log}%
\end{figure*}

To illustrate the superradiant character of the FW densities, we start with an
initial state $\Psi_{FW}(t_{0},x)=\left(  \varphi(t_{0},x),0\right)  $
representing a particle wavepacket with no antiparticle contribution.
$\varphi(t_{0},x)$ is chosen to have mean momentum and position given by
$x_{0}$ and $p_{0}$ respectively and is defined over a compact spatial support
of width $\Delta_{x}$; for numerical convenience, we choose
\begin{equation}
\begin{split}
	\varphi(t_{0},x)=(\theta(x-x_{0}+\Delta_{x}/2)-\theta(x-x_{0}-\Delta
_{x}/2)) \\
\times  \cos^{8}\left[  \frac{\pi}{\Delta_{x}}\left(  x-x_{0}\right)  \right]
e^{ip_{0}x} \,;%
\end{split}
 \label{eq:initwf}
\end{equation}
we will use natural units from now on, $\hbar=1$, $c=1$, $\lambda_{C}%
=\hbar/mc=1$ and set the charge $q=1$. We choose $p_{0}$ and $\Delta_{x}$
such that the wavepacket energy range corresponds to the Klein tunneling
regime, setting the potential height $V_{0}$ accordingly. For numerical
purposes, we replace each rectangular potential $V_{i}(x)$ of Eq.
\eqref{rectb} by
\begin{equation}
	V_{i}(x)=\frac{V_{0}}{2}\left[  \tanh(\epsilon(x-x_{i}+L/2))-\tanh(
	\epsilon(x-x_{i}-L/2))  \right]
\label{numv}
\end{equation} 
 with $\epsilon$ large.

A typical evolution across 7 supercritical potential barriers is shown in Fig.
\ref{fig:7potentialsEvolution}. The initial wavepacket is shown it Fig.
\ref{fig:7potentialsEvolution} (a), while panel (b) displays the charge
density as the wavepacket traverses the first barrier. While the FW
transformation ensures the initial and final states (\textquotedblleft
asymptotically\textquotedblright\ far from the barriers) are of pure particle
type ($\rho_{FW}$ is positive), elsewhere both components contribute to the
density ($\left\vert \varphi(t,x)\right\vert $ with a positive sign,
$\left\vert \chi(t,x)\right\vert $ with a negative sign indicative of the
antiparticle character) and $\rho_{FW}(t,x)$ can be locally positive or
negative, as seen in Fig. \ref{fig:7potentialsEvolution}(b). Note also the
superradiant character: both the positive and negative charges are locally
amplified, though the total charge must remain constant in time. Finally, Fig.
\ref{fig:7potentialsEvolution}(c) shows the density as the front of the
wavepacket exits the last potential barrier. A substantial fraction of the
wavepacket lies outside the light-cone (represented by the dashed vertical
lines), of the same order of magnitude as the initial wavepacket (the vertical
scale is the same across all plots). This is a very significant non-local
behavior due to the amplification of the charge density induced by the barriers.

The amplification of the density propagating outside the light-cone (OLC) is a combination of two ingredients: the superradiant aspect of supercritical potentials, leading to a local exponential charge increase,
and the instantaneous propagation known to characterize FW densities. 
Exponential charge increase is a well-known feature that
has been observed for standard (canonical) KG time-dependent densities
\cite{dombey,wavpackets-PRA-2021}
and in this case the wavepacket propagation is always causal.
Although FW wavefunctions generically propagate non-locally \cite{hegerfeldt-prl}, 
it is generally asserted \cite{ruijsenaars,pavsic,lorce2022,choi2024}, based on the behavior of field-free FW wavefunctions, that the OLC fraction of the
FW density is negligible, both in terms of charge intensity (exponentially decreasing) and spatial extent (on the order of 
the Compton wavelength), and therefore in practice not observable. 
However in the presence of superradiance, the OLC fraction
of the charge density is quickly amplified to becomes
of the same order of magnitude as the initial wavepacket. The effect of
superradiance can be seen in Fig. \ref{fig:density-log} (a), displaying the
dynamics shown in Fig. \ref{fig:7potentialsEvolution}(c) throughout the entire
spatial range (the reflected wavepacket is seen as well) and in logartihmic
scale.  The yellow dashed
line in Fig. \ref{fig:density-log} (a) represents the freely evolved FW density, obtained for the
same initial state $\Psi_{FW}(t_{0},x)=\left(  \varphi(t_{0},x),0\right)  $
with Eq. \eqref{eq:initwf} but evolving with no potential. The OLC fraction is
barely visible for free evolution despite the logarithmic scale.

To better contextualize the behavior of the FW density, Fig.\ref{fig:density-log} also shows, in panel (b), the standard Klein-Gordon density under identical initial conditions. It is computed using the same numerical method, but directly in the original representation with the Hamiltonian\eqref{eq:canonical-hamiltonian}. While both densities exhibit superradiant amplification due to tunneling through the supercritical barriers, only the FW density displays a fraction of charge propagating outside the light cone. In contrast, the canonical density remains strictly confined within the causal region, illustrating the fundamentally different propagation properties of the two formulations.
Note that although the initial compact density is the same in both cases, the standard representation necessarily involves a mixture of positive and negative energy components. This results in small negative values of the charge density even in the free region (visible on the left side of Fig.~\ref{fig:density-log}(b)), a well-known feature of relativistic wave propagation.

\begin{figure}[ptb]
	\centering
	\includegraphics[width=0.83\linewidth]{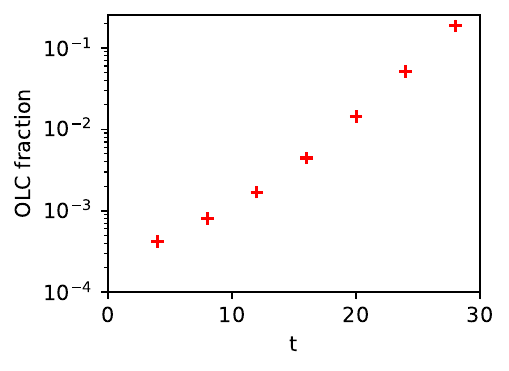} \caption{
		Fraction of the charge density lying outside the light cone at times
		$t_i$ at which the position $x_i$ of the light-cone emanating from the
		right-edge of the compact support shown in Fig.
		\ref{fig:7potentialsEvolution}(a) exits the $i$th 
		potential barrier ($i=1,...,7$). The OLC fraction is defined as 
		$\int_{x_{i}}^{\infty }dx \rho (t_i,x) $ (the total charge is normalized to 1).	
		Units and parameters for the
		potential barriers and the initial wavepacket are the same as in Fig.
		\ref{fig:7potentialsEvolution}.}%
	\label{fig:superradiance}%
\end{figure}

Let us further note that while with the present parameters, the non-causal
fraction of the wavefunction might lead to observational consequences if the
FW\ densities were physical, in principle we could envision additional
potential barriers and increase the value of $V_{0}$ to increase
dramatically the density propagating outside the light cone
(the OLC fraction is also enhanced if the wavepacket and barrier
widths are reduced, but in the present simulations $\Delta_{x}$ and $L$ are
already set to a few units of the Compton wavelength). This is supported (Fig.
\ref{fig:superradiance}) by computing the charge density as the light-cone reaches the
right edge of the $n$th barrier, confirming an exponential increase of the
OLC\ fraction as the number of tunneled barriers increases. Present
technologies are not yet able to produce supercritical fields leading to the
production of electron-positron pairs (e.g., \cite{LUXE}), let alone for the lightest
scalar meson (much heavier than an electron). Moreover, as known from quantum
field theory results \cite{grobe,Klein-tunneling-PRA-2022}, the second-quantized wavepacket density
is identical to the first quantized dynamics, but the dominant contribution to the overall particle
density near the potential edges comes from pair production, making it
harder to detect the wavepacket signal. However one could think of analogues
of superradiant Klein tunneling \cite{superradiant-equ}, in the spirit of the
well-known analogy of Klein tunneling for Dirac pseudo-particles in graphene
\cite{ref-graph}, that might lead to experimental observations.

To sum up, we have seen that in the particular instance of the Klein-Gordon
equation in the presence of supercritical potentials, Foldy-Wouthuysen
densities predict (in principle) observable properties that do not appear to
be acceptable on physical grounds, as they would lead to signaling. Indeed,
such issues that already appear but are claimed to be for all practical
purposes unobservable  in the field-free case \cite{ruijsenaars,pavsic,lorce2022,choi2024,rosenstein-usher,ruijgrok,eckstein,xabi}
 become amplified (in
intensity and spatial extent) by the superradiant character of Klein
tunneling.
These findings strongly suggest that the FW density, despite its formal derivation and classical appeal, cannot be regarded as a fundamentally correct charge or probability density in relativistic quantum mechanics. However, we emphasize that this does not invalidate the FW representation itself. As argued in previous works \cite{bialy2017,barnett2017,silenko2018,bliokh2017,viewpoint,bialy-comm,silenko-comm,choi2020,oppeneer2020}, particularly in studies on
relativistic electron beams obtained from the Dirac equation, the FW framework remains valuable. It provides quantum operators that align with classical observables such as position and spin, and it enables a transparent semiclassical analysis.
The consequence of the present
results on the choice of the position operator will be discussed elsewhere.

\appendix

\section{Obtaining the eigenvalue equation}
\label{supp-A}

The free FW transformation applied to the Klein-Gordon equation
\begin{equation}
	i\hbar\partial_{t}\Psi=\hat{\mathcal{H}}\Psi\label{timedep}%
\end{equation}
where
\begin{equation}
	\hat{\mathcal{H}}=mc^{2}\sigma_{3}+\frac{\hat{p}^{2}}{2m}(\sigma_{3}%
	+i\sigma_{2})+V(\hat{x})
\end{equation}
is the Klein-Gordon Hamiltonian [Eq. \eqref{eq:canonical-hamiltonian} of the main text] leads to
\begin{equation}
	i\hbar\frac{\partial}{\partial t}\ket{\Psi_{FW}(t)}=\left(  \left(  \hat
	{p}^{2}c^{2}+m^{2}c^{4}\right)  ^{1/2}\sigma_{3}+\hat{\mathcal{V}}%
	_{FW}\right)  \ket{\Psi_{FW}(t)} \,.
\end{equation}

The free part, giving the square-root term, can be found in relativistic quantum
mechanics textbooks (see e.g. \cite{greinerRQM}). The potential-dependent part is
computed straightforwardly as%
\begin{equation}
	\bra{p}\hat{\mathcal{V}}_{FW}\ket{\Psi}=\frac{1}{\sqrt{2\pi\hbar}}%
	\int_{-\infty}^{+\infty}\mathrm{d}p^\prime~\Tilde{V}(p-p^\prime)\mathcal{W}(p,p^\prime)\Psi(p^\prime)
\end{equation}
where%
\begin{equation}
	\Tilde{V}(p)=\frac{1}{\sqrt{2\pi\hbar}}\int_{-\infty}^{+\infty}%
	V(x)e^{-ipx/\hbar}\mathrm{d}x
\end{equation}
and%
\begin{equation}
	\mathcal{W}(p,p^\prime)=\mathcal{U}(p)\mathcal{U}^{-1}(p^\prime)=\frac{1}{2\sqrt{E_{p}E_{p^\prime}%
	}}\left[  (E_{p}+E_{p^\prime})\mathcal{I}+(E_{p}-E_{p^\prime})\sigma_{1}\right]
	\,.
	\label{eq:FW-pq}%
\end{equation}

Let us write $\mathcal{W}(p,p^\prime)$ in the form%
\begin{equation}
	\mathcal{W}(p,p^\prime)=%
	\begin{pmatrix}
		w^{(+)}(p,p^\prime) & w^{(-)}(p,p^\prime)\\
		w^{(-)}(p,p^\prime) & w^{(+)}(p,p^\prime)
	\end{pmatrix}
	\label{eq:FW-pq-matrix}%
\end{equation}
where%
\begin{equation}
	w^{(\pm)}(p,p^\prime)=\frac{E_{p}\pm E_{p^\prime}}{2\sqrt{E_{p}E_{p^\prime}}}\,.
\end{equation}

The eigenfunctions of $\mathcal{H}$ with eigenvalue $\epsilon_\lambda$ are found by separating the space and time variables as $\Psi_{\epsilon_\lambda}(t,p) = e^{-i\epsilon_\lambda t/\hbar}r_{\epsilon_\lambda}(p)$.
The equation for $r_{\epsilon_\lambda}(p)$ then takes the form
\begin{equation}
	\int_{-\infty}^{\infty}\mathrm{d}p^{\prime}\mathcal{K}(p,p^{\prime
	})r_{\epsilon_{\lambda}}(p^{\prime})=\epsilon_{\lambda}r_{\epsilon_{\lambda}%
	}(p)\label{eigen}%
\end{equation}
where%
\begin{equation}
	\mathcal{K}(p,p^\prime)=E_{p}\delta(p-p^\prime)\sigma_{3}+\frac{1}{\sqrt{2\pi\hbar}}%
	\Tilde{V}(p-p^\prime)\mathcal{W}(p,p^\prime).
\end{equation}

\section{Solving the eigenvalue equation}
\label{supp-B}

Let us denote again by  $\varphi_{\epsilon_{\lambda}}(p)$ and 
$\chi_{\epsilon_{\lambda}}(p)$ the two components of
\begin{equation}
	r_{\epsilon_{\lambda}}(p)=\left(
	\begin{array}
		[c]{c}%
		\varphi_{\epsilon_{\lambda}}(p)\\
		\chi_{\epsilon_{\lambda}}(p)
	\end{array}
	\right)
\end{equation}

We employ a discretization scheme particularly suitable for integral equations
\cite{num}, by which the integral equation
\eqref{eigen} is recast as an eigensystem problem%

\begin{equation}
	\epsilon_{n}r_{n}(p_{i})=\sum_{j=1}^{2N}\mathbf{K}_{i,j}r_{n}(p_{j})\Delta
	p\label{eq:discrete-integral}%
\end{equation}
where $r_{n}$ is the discretized $2N$-components vector
\begin{equation}
	r_{n}=%
	\begin{pmatrix}
		\varphi_{n}(p_{1})\\
		\vdots\\
		\varphi_{n}(p_{N})\\
		\chi_{n}(p_{1})\\
		\vdots\\
		\chi_{n}(p_{N})
	\end{pmatrix}
	\label{eq:vector-2}%
\end{equation}

We proceed similarly for the left eigenfunctions of the hamiltonian,%
\begin{equation}
	\epsilon_{n}^{\ast}l_{n}(p_{i})=\sum_{j=1}^{2N}\mathbf{K}_{i,j}^{\dagger}%
	l_{n}(p_{j})\Delta p\label{eq:discrete-integral2}%
\end{equation}
with a matrix $\mathbf{K}\in\mathcal{M}_{2N}(\mathbb{C})$ defined by blocks
$\mathbf{K}^{(a,b)}\in\mathcal{M}_{N}(\mathbb{C})$%
\begin{equation}
	\mathbf{K}=\left(
	\begin{array}
		[c]{@{}c|c}%
		\mathbf{K}^{(1,1)} & \mathbf{K}^{(1,2)}\\\hline
		\mathbf{K}^{(2,1)} & \mathbf{K}^{(2,2)}%
	\end{array}
	\right)  \label{eq:bloc-matrix}%
\end{equation}
where%
\begin{equation}
	\left\{
	\begin{array}
		[c]{ll}%
		\mathbf{K}_{i,j}^{(1,1)}=+\frac{E_{p_{i}}}{\Delta p}\delta_{i,j}+\frac
		{1}{\sqrt{2\pi\hbar}}\Tilde{V}(p_{i}-p_{j})w^{(+)}(p_{i},p_{j}) & \\
		\mathbf{K}_{i,j}^{(2,2)}=-\frac{E_{p_{i}}}{\Delta p}\delta_{i,j}+\frac
		{1}{\sqrt{2\pi\hbar}}\Tilde{V}(p_{i}-p_{j})w^{(+)}(p_{i},p_{j}) & \\
		\mathbf{K}_{i,j}^{(1,2)}=\mathbf{K}_{i,j}^{(2,1)}=+\frac{1}{\sqrt{2\pi\hbar}%
		}\Tilde{V}(p_{i}-p_{j})w^{(-)}(p_{i},p_{j}) &
	\end{array}
	\,.\right.
\end{equation}
$\Delta p=2\pi/(x_{max}-x_{min})$ is the step in the discretized momentum
space and $N$ the number of discretization points. Hence for a finite position
interval $[x_{min},x_{max}]$, we have a step $\Delta x$ with $x_{1}=x_{min}$,
$x_{j}=x_{min}+j\Delta x$ and $x_{N}=x_{max}$. Our Python implementation relies on the \texttt{eig} function from SciPy's linear algebra module, which conveniently provides both left and right eigenvectors, along with their associated eigenvalue.

\section{The role of complex eigenvalues}
\label{supp-C}

The pseudo-Hermitian character of the Klein-Gordon equation is evident from the 
definition of the  scalar product, defined by

\begin{equation}
	\label{product}
	\braket{\Psi_{1} | \Psi_{2}}_{\sigma_3} =\int
	dx\Psi_{1}^{\dagger}(x)\sigma_{3}\Psi_{2}(x)
\end{equation}
with the ubiquitous presence of the $\sigma_{3}$ term. 
A pseudo-Hermitian operator is neither self-adjoint nor  normal (satisfying
$[\mathcal{H},\mathcal{H}^{\dagger}]=0$) \cite{mostafazadeh}. A pseudo-Hermitian Hamiltonian 
such as $\mathcal{H}_{FW}$ admits a biorthogonal expansion on a basis consisting of its ``right'' and ``left'' eigenvectors defined 
by
\begin{equation}
	\left\{
	\begin{array}
		[c]{ll}%
		\hat{\mathcal{H}}_{FW}\ket{r_{\epsilon_\lambda}}=\epsilon_{\lambda
		}\ket{r_{\epsilon_\lambda}}\\
		\hat{\mathcal{H}}_{FW}^{\dagger}\ket{l_{\epsilon_\lambda}}=\epsilon_{\lambda}%
		^{\ast}\ket{l_{\epsilon_\lambda}}
	\end{array}
	\right.  \label{eq:eigen-rl}%
\end{equation}

Hence, while Hermicity implies that the eigenvalues are real, 
in the pseudo-Hermitian case the eigenvalues are not
necessarily real but come in pairs of complex conjugates \cite{mostafazadeh}. The
general relation between left and right eigenvectors in our problem is%

\begin{equation}
	\label{relation}
	\ket{l_{\epsilon_\lambda^*}} = \sigma_{3} \ket{r_{\epsilon_\lambda}}
\end{equation}

It is worth noting that the use of the biorthonormal basis in the decomposition of our initial wavefunction [Eq. \eqref{eq:psi-FW}] circumvents the need for a more involved formulation in terms of pseudo-inner products using $\sigma_{3}$.

It should also be noted that eigenfunctions associated with non-real eigenvalues of the Hamiltonian
have an amplitude exponentially increasing or decreasing over time. Such terms 
account for superradiance (exponential local increase of the charge).
However the total charge $q$ of a state $\ket{\Psi(t)}=\hat{\mathcal{T}}(t-t_0)\ket{\Psi(t_0)}$ is still conserved since it is given by

\begin{equation}
	\int \mathrm{d}x \rho(t,x)
	= \int \mathrm{d}x q \Psi^\dagger(t,x) \sigma_3 \Psi(t,x)\\
	= q \braket{\Psi(t) | \Psi(t)}_{\sigma_{3}}\,,
\end{equation}

\noindent
and the evolution operator

\begin{equation}
	\hat{\mathcal{T}}(t-t_0) = \sum_{\lambda} e^{-i\epsilon_{\lambda}(t-t_0)/\hbar} \ket{r_{\epsilon_\lambda}} \bra{l_{\epsilon_\lambda}}
\end{equation}

\noindent
is unitary with respect to this pseudo-inner product \eqref{product} so that 

\begin{equation}
	\braket{\Psi(t) | \Psi(t)}_{\sigma_{3}}
	= \braket{\Psi(t_0) | \Psi(t_0)}_{\sigma_{3}}
	= 1\,.
\end{equation}

\noindent
for an initially normalized state.
This can readily be verified using property \eqref{relation}.


\begin{thebibliography}{99}                                                                                     
	\bibitem{review-eb} S.M. Lloyd, M. Babiker, G. Thirunavukkarasu, and J. Yuan, Electron vortices: Beams with orbital angular momentum, Rev. Mod. Phys. 89, 035004 (2017).
	\bibitem{bialy2017}	I. Bialynicki-Birula and S. Bialynicka-Birula, Relativistic Electron Wave Packets Carrying Angular Momentum, Phys. Rev. Lett. 118, 114801 (2017).
	\bibitem {FW}  L. L. Foldy and S. A. Wouthuysen, On the Dirac Theory of Spin 1/2 Particles and Its Non-Relativistic Limit, Phys. Rev. 78, 29 (1950).
	\bibitem {greinerRQM}Greiner, W. Relativistic Quantum Mechanics. (Springer Berlin Heidelberg,1990).
	
	\bibitem {barnett2017}S. M. Barnett, Relativistic Electron Vortices, Phys. Rev. Lett.  118, 114802 (2017).
	\bibitem {silenko2018}A. J. Silenko, P. Zhang and L. Zou, Relativistic Quantum Dynamics of Twisted Electron Beams in Arbitrary Electric and Magnetic Fields, Phys. Rev. Lett.  118, 114802 (2018).
	\bibitem {bliokh2017} K. Y. Bliokh, M. R. Dennis and F. Nori, Position, Spin, and Orbital Angular Momentum of a Relativistic Electron, Phys. Rev. A , 023622 (2017).
	\bibitem {viewpoint} H.\ Larocque and E.\ Karimi, A New Twist on Relativistic Electron Vortices,  Physics 10, 26 (2017).
	\bibitem {bialy-comm} I. Bialynicki-Birula and S. Bialynicka-Birula, Comment on “Relativistic Quantum Dynamics of Twisted Electron Beams in Arbitrary Electric and Magnetic Fields”,  Phys. Rev. Lett.  122, 159301 (2019).
	\bibitem {silenko-comm}A. J. Silenko, P. Zhang and L. Zou, Phys. Rev. Lett. 122, 159302 (2019).
	\bibitem {choi2020}Y. D. Han, T. Choi and S. Y. Cho, Singularity of a relativistic vortex beam and proper relativistic observables, Sci Rep 10, 7417 (2020).
	\bibitem {oppeneer2020} R. Mondal and P. M. Oppeneer, Dynamics of the relativistic electron spin in an electromagnetic field, J. Phys.: Condens. Matter 32  455802 (2020).
	\bibitem {pryce}M. H. L. Pryce, The mass-centre in the restricted theory of relativity and its connection with the quantum theory of elementary particles, Proc. R. Soc. London A 195, 62 (1948).
	\bibitem {newton-wigner}    T. D. Newton and E. P. Wigner, Localized states for elementary systems, Rev. Mod. Phys. 21, 400 (1949).
	
	\bibitem{kalnay1971} A. J. Kalnay, The Localization Problem, in \textit{Problems in the Foundations of Physics}, edited by M. Bunge, Studies in the Foundations, Methodology and Philosophy of Science Vol. IV (Springer-Verlag, Berlin, 1971).
	
	\bibitem {case}  K. M. Case,  Some Generalizations of the Foldy-Wouthuysen Transformation, Phys. Rev. 95, 1323 (1954).
	\bibitem {silenko2020}L. Zou, P.  Zhang and A. J. Silenko, Position and Spin in Relativistic Quantum Mechanics, Phys. Rev. A 101, 032117 (2020)
	\bibitem {hegerfeldt-prl}G. C. Hegerfeldt, Violation of Causality in Relativistic Quantum Theory ?, Phys. Rev. Lett. 54 2395 (1985).
	\bibitem {ruijsenaars} S.N.M Ruijsenaars, On Newton-Wigner Localization and Superluminal Propagation Speeds, Ann. Phys. 137, 33 (1981).
	\bibitem {pavsic} M. Pavsic, Localized States in Quantum Field Theory, Localized states in quantum field theory, Adv. Appl. Clifford Algebras 28 89 (2018).
	
	\bibitem {lorce2022}Y. Chen and C. Lorc\'{e}, Pion and nucleon relativistic electromagnetic four-current distributions, Phys. Rev. D 106, 116024 (2022).
	
	\bibitem {choi2024}T.\ Choi, Lorentz-covariance of Position Operator and its Eigenstates, Int. J. Theor. Phys. 63, 10 (2024).
	\bibitem {rosenstein-usher}B. Rosenstein and M. Usher,Explicit illustration of causality violation: noncausal relativistic wave-packet evolution Phys. Rev. D 36 2381 (1987).
	
	\bibitem {ruijgrok}T.\ W.\ Ruijgrok, On Localisation in Relativistic Quantum Mechanics, Lect. Notes Phys. 539, 52 (2000).
	\bibitem {eckstein}M. Eckstein and T. Miller, Causal evolution of wave packets. Phys. Rev. A 2017, 95, 032106 (2017).
	
	\bibitem{xabi} X. Gutierrez de la Cal X and A. Matzkin, Beyond the light-cone propagation of relativistic wavefunctions: numerical results, Dynamics 3 60 (2023).
	
	\bibitem{manogue} C. A. Manogue, The Klein paradox and superradiance, Ann. Phys. 181, 261 (1988).	
	\bibitem{dombey} N. Dombey and A. Calogeracos, Seventy years of the Klein paradox, Phys. Rep. 315, 41 (1999).
	\bibitem{wavpackets-PRA-2021}M. Alkhateeb, X. Gutierrez de la Cal, M. Pons, D. Sokolovski, and A. Matzkin, Relativistic time-dependent quantum dynamics across supercritical barriers for Klein–Gordon and Dirac particles. Phys. Rev. A  103, 042203 (2021). 
	\bibitem{bauke}M. Ruf, H. Bauke, and C. H. Keitel, J. Comput. Phys. 228, 9092 (2009).
	
	\bibitem{mostafazadeh}A. Mostafazadeh, Pseudo-Hermitian Representation of Quantum Mechanics,  Int. J. Geom. Meth. Mod. Phys. 7  1191 (2010).
	
	\bibitem {hermanfeshbachElementaryRelativisticWave1958}H. Feshbach and F. Villars, Elementary Relativistic Wave Mechanics of Spin 0 and Spin 1/2 Particles, Rev. Mod. Phys. 30, 24 (1958).
	
	\bibitem{num}A. D. Poljanin and  Andrej D. and A. V. Man{\v z}irov, \textit{Handbook of Integral Equations}(CRC Press, Boca-Raton, USA, 1998)
	
	\bibitem{LUXE}H. Abramowicz et al., Conceptual design report for the LUXE experiment, Eur. Phys. J. Spec. Top. 230, 2445 (2021).
	
	\bibitem{grobe}P. Krekora, Q. Su, and R. Grobe, Klein Paradox in Spatial and Temporal Resolution, Phys. Rev. Lett. 92, 040406 (2004).
	
	\bibitem{Klein-tunneling-PRA-2022}M. Alkhateeb and A. Matzkin, Space-time-resolved quantum field approach to Klein-tunneling dynamics across a finite barrier, Phys. Rev. A 106, L060202 (2022).
	
	\bibitem{superradiant-equ} A. Delhom et al., Entanglement from superradiance and rotating quantum fluids of light, Phys. Rev. D 109, 105024 (2024).
	
	\bibitem{ref-graph}A. H. Castro Neto, F. Guinea, N. M. R. Peres, K. S. Novoselov, and A. K. Geim, The electronic properties of graphene, Rev. Mod. Phys. 81, 109 (2009)
	
	
	
\end{thebibliography}
\end{document}